\documentclass[usenatbib,longtabl]{mn2e}

\title[W UMa contact binaries]
{The energy transfer and its effects on the secondaries in W Ursae
Majoris-type contact binaries}

\author[Jiang et al.]{Dengkai Jiang$^{1,2}$\thanks{E-mail:
jiangdengkai@hotmail.com},
Zhanwen Han$^{1}$, Tianyu Jiang$^{1,2}$, and Lifang Li$^{1}$\\
$^{1}$National Astronomical Observatories, Yunnan Observatory,
Chinese Academy of Sciences, P.O. Box 110,
Kunming,\\
\ \ \ \  \ \ \ \ \ \ \ \ \ \ \ \ \ \ \ \ \ \ \ \ \ \ \ \ \ \ \ \ \
\ \ \ \  Yunnan Province, 650011, P.R. China\\
$^{2}$Graduate University of Chinese Academy Sciences, Beijing,
100039}
\begin{document}
\input ctextemp_psfig.sty
\date{Accepted .... Received .....; in original form ....}

\pagerange{\pageref{firstpage}--\pageref{lastpage}} \pubyear{2008}

\maketitle

\label{firstpage}

\begin{abstract}
Based on the physical parameters of 133 W Ursae Majoris (W
UMa)-type contact binaries, the energy transfer and its effects on
the secondary in W UMa contact binaries are investigated.
Relations are given between the mass ratio ($q$) for W UMa contact
binaries and the relative energy transfer rates, i.e. $U_1$, the
ratio of the transferred luminosity to the surface luminosity of
the primary and $U_2$, the ratio of the transferred luminosity to
the nuclear luminosity of the secondary. {\bf The theoretical
curves($U_1$ $vs$ $q$ and $U_2$ $vs$ $q$) are derived based on the
various assumptions that the two components in each W UMa system
are nearly identical in effective temperature, they just fill
their inner Roche lobes, and the primaries are ZAMS stars.
Although these curves can reflect the distribution of $U_1$  $vs$
$q$ and $U_2$ $vs$ $q$, some observational systems are
significantly deviated from these curves. It is mainly resulted
from the difference in the effective temperatures of the
components in W UMa systems.}The radius and the density of the
secondary are related to the relative energy transfer rate $U_2$:
the higher is $U_2$, the greater is the expansion and the lower is
the density of the secondaries in W UMa systems. In addition, it
is found that the temperature difference of W UMa binary
components is correlated with the relative energy transfer rate
$U_1$ and decreases with increasing $U_1$. This might suggest that
there is a thermal coupling between two components in W UMa
contact binaries, and that the classification of W UMa contact
binaries into A- or W-types depends on the energy transfer from
the primary to the secondary. The temperature difference of W UMa
binary components is poorly correlated with the mass of the
primary. This suggests that the properties of the common envelope
of W UMa contact binaries might not have a significant effect on
the energy transfer between two components.
\end{abstract}


\section{Introduction}

W UMa contact binaries are very common eclipsing variables in which
the eclipsing light curves have nearly equal minima. \citet{bin70}
classified W UMa contact binaries into A- or W-type on the basis of
their light curves. The secondaries of W UMa contact binaries have
an unusual mass-luminosity relationship which was first recognized
by \citet{Struve 1948}. \citet{Lucy 1968} proposed that it is caused
by the energy transfer from the primary (the more massive component)
to the secondary (the less massive component) within a convective
envelope. But the mechanism causing energy transfer between the two
components of W UMa contact binaries and the effect of the energy
transfer on the components are not clear.

W UMa contact binaries are an important class of eclipsing variables
in several respects. In studies of Galactic structure, W UMa contact
binaries play an important role because they have high spatial
frequency of occurrence, ease of detection, and provide a standard
candle for distance determinations \citep{Rucinski 1997}. More
importantly, W UMa contact binaries are interesting objects due to
the mass and energy transfer between two components. Understanding
the energy and mass transfer in the common envelope is necessary to
develop a correct theory of the structure and evolution of W UMa
systems. Therefore, the investigation of the mechanism causing
energy transfer is a core problem for understanding the structure
and evolution of W UMa contact binaries \citep{Webbink 2003}. The
energy transfer in W UMa systems has been investigated by many
authors \citep{Mochnacki 1981, Smith 1984, Kaluzny 1985, Hilditch
1988}. \citet{Mochnacki 1981} calculated the relative energy
transfer rate of W UMa contact binaries using the normal
mass-luminosity relation for independent stars and found that the
relative energy transfer rate depends only on the mass ratio of W
UMa contact binaries. \citet{Wang 1994} found that the relative
energy transfer rate increases with increasing mass ratio
continuously based on 22 contact binaries. \citet{Liu 2000} found
that the energy transfer rate depends not only on the mass ratio but
also on the evolutionary degree of the primary. By studying a
catalogue data of 159 systems, \citet{Csizmadia 2004} found that the
energy transfer rate is a function of the mass and luminosity ratio.

The treatment of the energy transfer is very important to construct
theoretical models of W UMa contact binaries. Although it seems
probable that the energy transfer occurs in the common envelope of W
UMa systems, it is not clear at present where and how the transfer
is taking placed in the common envelope. The structure and evolution
of W UMa contact binaries have been investigated by several authors
in recent years \citep{Kahler 2002a, Kahler 2002b, Li 2004, Li 2005,
Yakut 2005}. It is found that loss of contact is avoided if the
energy transfer is assumed to be sufficiently effective
\citep{Kahler 2002a, Kahler 2002b}. \citet{Li 2004} discussed the
region of energy transfer in the common envelope of W UMa contact
binaries, and showed that the energy transfer may take place in the
radiative region of the common envelope. \citet{Yakut 2005}
suggested that the mechanism of the energy transfer may be
differential rotation which has been observed by helioseismology in
the solar convection zone \citep{Schou 1998}.

The energy transfer from the primary to the secondary in W UMa
contact binaries will restructure the secondary and make it
oversized and overluminous for its mass \citep{Webbink 2003}. The
radii of the secondaries of W UMa contact binaries (including A- and
W-types) are obviously altered from those of zero-age main sequence
(ZAMS) stars \citep{Yakut 2005,Li 2008}. This might be the result of
the energy transfer from the primary to the secondary \citep{Yang
2001, Li 2008}. The temperatures and densities of the secondaries of
W UMa contact binaries are also affected by the energy transfer.
\citet{Hazlehurst 1977} calculated the response functions which
describe the effect of energy transfer on the radii and effective
temperatures of the stars. But the relation between the energy
transfer and the reaction of the secondaries of W UMa systems is not
completely understood.

In this paper, the physical parameters of 133 W UMa contact binaries
have been collected from the literature. Using these data, the
energy transfer and its effect on the secondaries in W UMa systems
are investigated.

\section{Energy transfer in W UMa systems}

\begin{table*}
\begin{footnotesize}
Table~1.\hspace{4pt} The new or recently obtained physical parameters of contact binaries.\\
\begin{minipage}{17cm}
\begin{tabular}{l|cccccccccccc}
\hline\hline\
{Stars}&{Type}&{$P$}&{$M_{1}$}&{$M_{2}$}&$R_{1}$& $R_{2}$&{$L_{1}$}&{$L_{2}$}&{$T_{1}$}&{$T_{2}$}&{$q_{_{\rm ph}}$}&{References}\\
&&{(days)}&{($M_{\rm \odot}$)}& {($M_{\rm \odot}$)}&{($R_{\rm
\odot}$)}&{($R_{\rm \odot}$)}&{($L_{\rm \odot}$)}& {($L_{\rm \odot}$)}& {(K)}& {(K)}&&\\
\hline

VW LMi&W&0.4775&1.67&0.70&1.709&1.208&5.31&2.80&6700&6792&0.416&1\\
BX Dra&A&0.5790&2.19&0.63&2.13&1.26&9.8&2.5&7000&6446&0.289&1\\
DN Bootis&A&0.4476&1.428&0.148&1.710&0.67&3.750&0.560&6095&6071&0.103&2\\
\hline
\end{tabular}
\end{minipage}
\end{footnotesize}\\
{Columns: Stars-GCVS name of star; $P$-orbital period; $M_1$-mass of
the primary; $M_2$-mass of the secondary; $L_1$-luminosity of
the\\
primary; $L_2$-luminosity of the secondary.$R_1$-radius of the
primary;$R_2$-radius of the secondary;
$T_1$-effective temperature of the\\
primary; $T_2$- effective temperature of the secondary\\
References in Table 1: (1) S\'{a}nchez-Bajo et al. 2007; (2)
\c{S}enavc{\i} et al. 2008}
\end{table*}

The physical parameters of 130 W UMa contact binaries were collected
from the compilations of \citet{Yakut 2005, Awadalla 2005, Maceroni
1996},{\bf and \citet{ Li 2008}.  The temperatures used in our paper
are also taken from the same sources. In addition, the new or
recently obtained physical parameters of 3 W UMa contact binaries
were collected from other sources (listed in Table 1).} Based on
these data, the evolutionary properties of W UMa systems are
analyzed. The observations suggest that the secondaries of W UMa
systems are overluminous and oversized \citep{Yang 2001, Webbink
2003, Stepien 2006, Li 2008}. Up to now, although the physical cause
for the over-luminosity and over-volume of the secondaries of W UMa
systems is not known, the similar appearance for the secondaries of
W UMa systems suggests that a common mechanism would produce them.
Two possible hypotheses have been proposed to explain the
over-luminosity and over-volume of the secondaries in W UMa systems,
either (a) energy transfer between the two components \citep{Lucy
1968, Webbink 2003, Li 2008}, or (b) W UMa systems with a more
evolutionarily advanced secondary due to the reversal of the mass
ratio \citep{Stepien 2006}.

\begin{figure}
\centerline{\psfig{figure=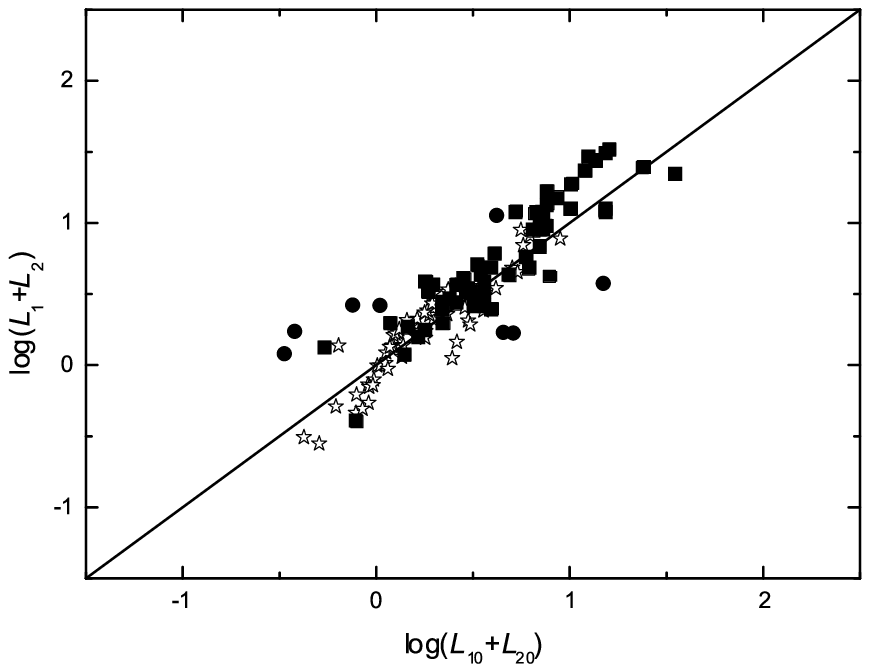,width=7.5cm}} \caption{Relation
between ${\rm log}(L_1+L_2)$ and ${\rm log}(L_{1, \rm nuc}+L_{2, \rm
nuc})$ for A-, and W-type W UMa contact binaries. Full squares and
open stars represent A-types and W-Types, respectively. The solid
line represents ${\rm log}(L_1+L_2)={\rm log}(L_{10}+L_{20})$, and
full circles represent the systems significantly deviating the solid
line.} \label{fig1}
\end{figure}

The distribution of ${\rm log}(L_1+L_2)$ $vs$ ${\rm
log}(L_{10}+L_{20})$ of our sample is presented in Figure 1.
$L_{1,2}$ are the surface luminosity of the primary and the
secondary. $L_{10}$ and $L_{20}$ are the nuclear luminosity of the
primary and the secondary. According to {\bf \citet{Demircan 1991}},
the nuclear luminosity of each component in W UMa contact binary can
be expressed as
\begin{equation}
L_0 \cong 1.03M^{3.42} \;\;\;\;  0.1\le M \le 120
\end{equation}
where $M$ is the mass of the component in solar units. The solid
line represents the case that the total surface luminosity is equal
to the total nuclear luminosity. It is seen in Figure 1 that most W
UMa contact binaries are located on the solid line within
observational errors. This indicates that the total observational
luminosity of W UMa contact binaries is nearly equal to the total
nuclear luminosity,
 implying an important evidence that energy is indeed
transferred from the primary to the secondary in W UMa systems
\citep{Mochnacki 1981, Webbink 2003}. Meanwhile, it also suggests
that the over-luminosity of secondaries in W UMa systems is
attributed to the energy transfer between two components. It is
also noted that at least {\bf 8} data points deviate strongly from
the line and the corresponding systems with $|{\rm
log}(L_1+L_2)$$-$${\rm log}(L_{10}+L_{20})|>0.4$ are listed in
table 2. {\bf In these systems, three high-mass systems (AC Boo,
ET Leo and V899 Her) with total mass larger than $2M_{\rm \odot}$
are below the solid line. This kind of deviation is not impossibly
caused by the evolved components in these systems and probably
caused by the inaccurate spectroscopic solution because of the
presence of the additional companions \citep{Pribulla 2008}. The
other four low-mass systems (BH Cas, DX Tuc, TZ Boo and XY Boo)
with total mass less than $1.4M_{\rm \odot}$ and one high-mass
system (V1073 Cyg) are above the solid line.} This kind of
deviation can be caused not only by the presence of the additional
companions, but also by the evolved components contained in these
systems. In fact, \citet{Bilir 2005} found that low-mass W UMa
contact binaries show the larger velocity dispersions than
high-mass W UMa contact binaries and low-mass W UMa systems have a
larger mean kinematic age than high-mass W UMa systems, implying
that some components of W UMa contact binaries with very low total
masses are significantly evolved. On the other hand,
\citet{Pribulla 2006} found that up to 59 percent of W UMa contact
binaries have companions. Meanwhile, \citet{Pribulla 2008} show
that TZ Boo is quadruple system and its spectra is contaminated by
third and fourth bodies. Therefore, the effect of the companions
on the spectroscopic solution of these systems might be one of the
reasons which lead these systems to deviate significantly from the
solid line.

\begin{table*}
\begin{footnotesize}
Table~2.\hspace{4pt} Physical parameters of most discrepant contact binaries.\\
\begin{minipage}{17cm}
\begin{tabular}{l|cccccccccccc}
\hline\hline\
{Stars}&{Type}&{$P$}&{$M_{1}$}&{$M_{2}$}&$R_{1}$& $R_{2}$&{$L_{1}$}&{$L_{2}$}&{$T_{1}$}&{$T_{2}$}&{$q_{_{\rm ph}}$}&{References}\\
&&{(days)}&{($M_{\rm \odot}$)}& {($M_{\rm \odot}$)}&{($R_{\rm
\odot}$)}&{($R_{\rm \odot}$)}&{($L_{\rm \odot}$)}& {($L_{\rm \odot}$)}& {(K)}& {(K)}&&\\
\hline
AC Boo&W&0.3524&1.534&0.476&1.314&0.572&1.427&0.269&5530&5520&0.31&1\\
V1073 Cyg&A&0.7859&1.498&0.479&2.154&1.318&8.263&3.020&6700&6661&0.320&1\\
ET Leo&W&0.3465&1.586&0.542&1.359&0.835&1.115&0.564&5112&5500&0.342&2\\
V899 Her&A&0.4212&2.1&1.19&1.57&1.22&2.32&1.44&5700&5677&0.566&3\\
TZ Boo&A&0.2976&0.72&0.11&0.97&0.43&1.02&0.18&5890&5754&0.153&4\\
BH Cas&W&0.4059&0.73&0.35&1.09&0.78&1.01&0.72&5550&6000&0.475&5\\
XY Boo&A&0.3706&0.912&0.169&1.230&0.607&2.138&0.515&6324&6307&0.1855&6\\
DX Tuc& A&0.3771&1.00&0.30&1.20&0.71&1.97&0.66&6250&6182&0.29&7\\
\hline
\end{tabular}
\end{minipage}
\end{footnotesize}\\
{Columns: Stars-GCVS name of star; $P$-orbital period; $M_1$-mass of
the primary; $M_2$-mass of the secondary; $L_1$-luminosity of
the\\
primary; $L_2$-luminosity of the secondary.$R_1$-radius of the
primary;$R_2$-radius of the secondary;
$T_1$-effective temperature of the\\
primary; $T_2$- effective temperature of the secondary\\
References in Table 1: (1) Awadalla \& Hanna 2005; (2) Gazeas et al.
2006; (3) \"Ozdemir et al. 2002; (4) Yakut et al. 2005; (5) Zo{\l}a
et al. 2001 ; (6) Yang et al. 2005; (7) Szalai et al. 2007}
\end{table*}

\citet{Mochnacki 1981} defined a relative energy transfer rate which
is the ratio of the transferred luminosity to the surface luminosity
of the primary. {\bf Based on the assumption that the primaries are
ZAMS, the relative energy transfer rate $U_1$ can be written as
\begin{equation}
U_1 =\frac{{\rm d}L}{L_1}=
\frac{\frac{L_2}{L_1}-\frac{L_{20}}{L_{10}}}{1+\frac{L_{20}}{L_{10}}},
\end{equation}
where $L_{1,2}$ are the surface luminosity of the primary and the
secondary; $L_{10}$ and $L_{20}$ are their nuclear luminosities,
respectively \citep{Mochnacki 1981, Wang 1994}. We can give
another relative energy transfer rate as
\begin{equation}
U_2 = {\rm log}(\frac{{\rm d}L}{L_{20}})= {\rm
log}(\frac{\frac{L_2}{L_1}-\frac{L_{20}}{L_{10}}}{\frac{L_{20}}{L_{10}}(1+\frac{L_{2}}{L_{1}})}).
\end{equation}
Using equation (1), we have
\begin{equation}
U_1= \frac{r^2t^4-q^{\alpha}}{1+q^{\alpha}},
\end{equation}
\begin{equation}
U_2 = {\rm log}(\frac{r^2t^4-q^{\alpha}}{q^{\alpha}(1+r^2t^4)}).
\end{equation}
where $r=R_2/R_1$,$t=T_2/T_1$, $q$ is mass ratio and $\alpha$ is the
exponent of mass-luminosity relation and it is equal to 3.42. If it
is assumed that the components in each W UMa system are identical in
the effective temperature, and that the components of each W UMa
contact binary just fill the inner Roche lobes (i.e.
$R_2/R_1=q^{0.46}$), equation (4) and equation (5) can be written as
\begin{equation}
U_1=\frac{q^{0.92}-q^{3.42}}{1+q^{3.42}}.
\end{equation}
\begin{equation}
U_2 = {\rm log}(\frac{q^{0.92}-q^{3.42}}{q^{3.42}(1+q^{0.92})}).
\end{equation}
The theoretical curves and the observational data are shown in
Figure 2 and Figure 3 with a solid line and the open stars
(W-subtypes) or the solid squares (A-subtypes), respectively. It is
seen in Figure 2 and Figure 3 that although the theoretical curves
can reflect the distribution of the $U_1$ $vs$ $q$ and $U_2$ $vs$
$q$ of the observed data, the observational points are largely
scattered and some observed systems are significantly deviated from
the solid line. The deviation might be caused by the applicability
of the basic assumptions.

In order to find the applicability of the basic assumptions, we take
$r$, $t$, and $\alpha$ to be different values. At first, we must
inspect the applicability of a basic assumption that the components
of each W UMa system just fill their inner Roche lobes (i.e.
$R_2/R_1=q^{0.46}$). In fact, most observed systems are over-contact
binaries, and they should not satisfy this relation. The relation
between the logarithms of the radius ratio ($R_2/R_1$) and the
logarithms of the mass ratio $q$ of the observed systems is shown in
Figure 4. As seen from Figure 4, the logarithm of the radius ratio
is almost linearly changed with the logarithms of the mass ratio. A
least-squares solution leads to the following relation,
\begin{equation}
{\rm log}(R_2/R_1)=0.431(6){\rm log}q-0.007(3).
\end{equation}
According to equation (8), $r=R_2/R_1=0.984q^{0.43}$, which is
indeed different from Roche approximation relation $r=q^{0.46}$.
These relations are also plotted in Figure 4 with a dashed line and
a solid line, respectively. Using the relation,
$r=R_2/R_1=0.984q^{0.43}$, of the observed systems, equation (4) and
equation (5) are shown in Figure 2 and Figure 3 with a dot-dashed
line, respectively. As seen from Figure 2 and Figure 3, the
dot-dashed lines are very similar to the solid ones. Therefore, the
assumption that the components of each W UMa systems just fill their
inner Roche lobes is acceptable. Secondly, the different values of
$\alpha$ are adopted by the different investigators (4.0,
\citet{Wang 1994, Mochnacki 1981}, 4.6, \citet{Csizmadia 2004}). If
$\alpha=4.6$ (the largest value), the relative distributions between
the observed systems and the theoretical curves are similar to those
shown by Figures 2 and 3, although the observed points and
theoretical curves have been shifted since they are shifted with the
same direction. This suggests that the exponent of the
mass-luminosity relation has little effects on the relative
distribution between the observed systems and the theoretical
curves. So, the assumption that the components in W UMa systems are
ZAMS is also acceptable. Finally, we inspect the assumption of the
two components with equal temperature in each W UMa system. The
distribution of $t$ for the observed W UMa systems is located in a
region from 0.88 to 1.097. We take $t$ to be 0.9 and 1.1.  We plot
the resulting curves in Figure 2 and Figure 3 with two dashed lines.
As seen from Figure 2 and Figure 3, these two dashed lines can cover
the large scattering of the observed systems. Therefore, the
deviation caused by unequal effective temperatures of the components
of W UMa systems are larger than those caused by other two
assumptions. This suggests that the equal-temperature assumption for
the components of W UMa systems is the most unreasonable one in
three basic simplifying assumptions.}

\begin{figure}
\centerline{\psfig{figure=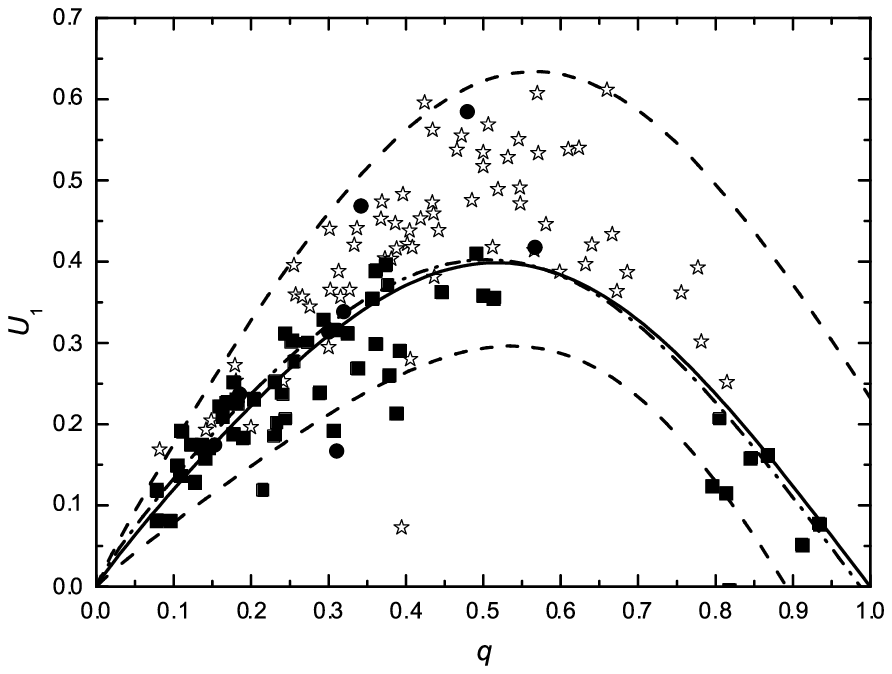,width=7.5cm}} \caption{The
relative energy transfer rate $U_1$ as a function of mass ratio of W
UMa contact binaries. The symbols are the same as Figure 1 and these
curves are derived from the assumptions(see the text).} \label{fig2}
\end{figure}

\begin{figure}
\centerline{\psfig{figure=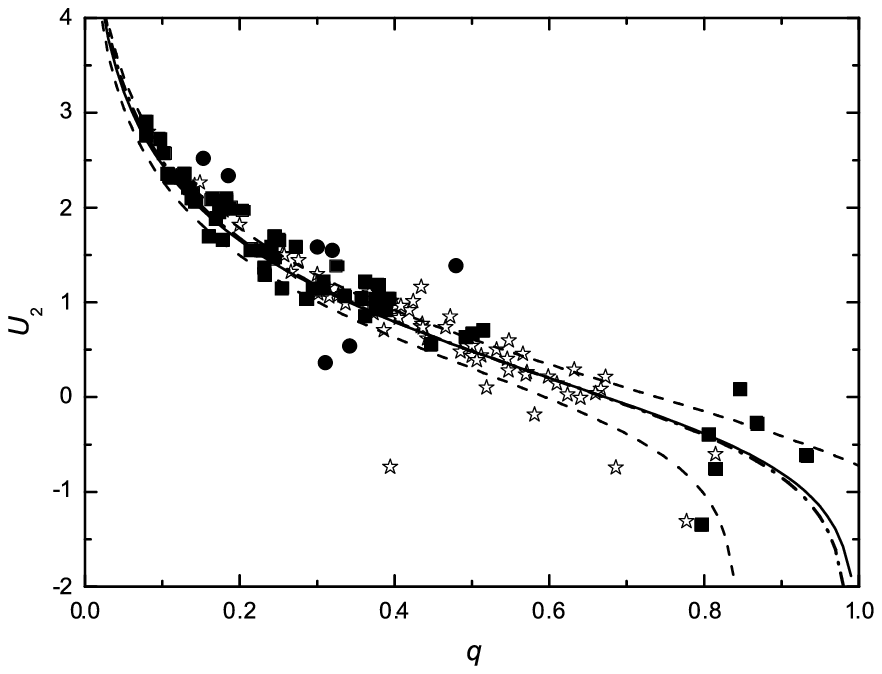,width=7.5cm}} \caption{The
relative energy transfer rate $U_2$ as a function of mass ratio of W
UMa contact binaries. The symbols are the same as Figure 1 and these
curves are derived from the assumptions(see the text).} \label{fig5}
\end{figure}

\begin{figure}
\centerline{\psfig{figure=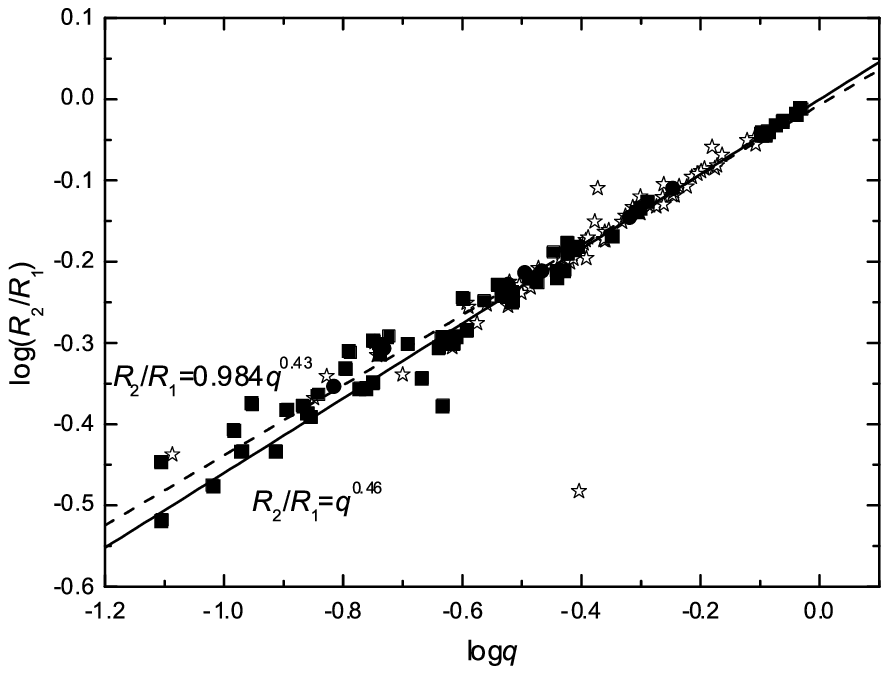,width=7.5cm}} \caption{Relation
between ${\rm log}(R_2/R_1)$ and ${\rm log}(q)$ for A-, and W-type W
UMa contact binaries. Full squares and open stars represent A-types
and W-Types, respectively. The solid line represents ${\rm
log}(R_2/R_1)=0.43{\rm log}(q)$. The dash line represents the fitted
curve and the symbols are the same as Figure 1.} \label{fig3}
\end{figure}

\begin{figure}
\centerline{\psfig{figure=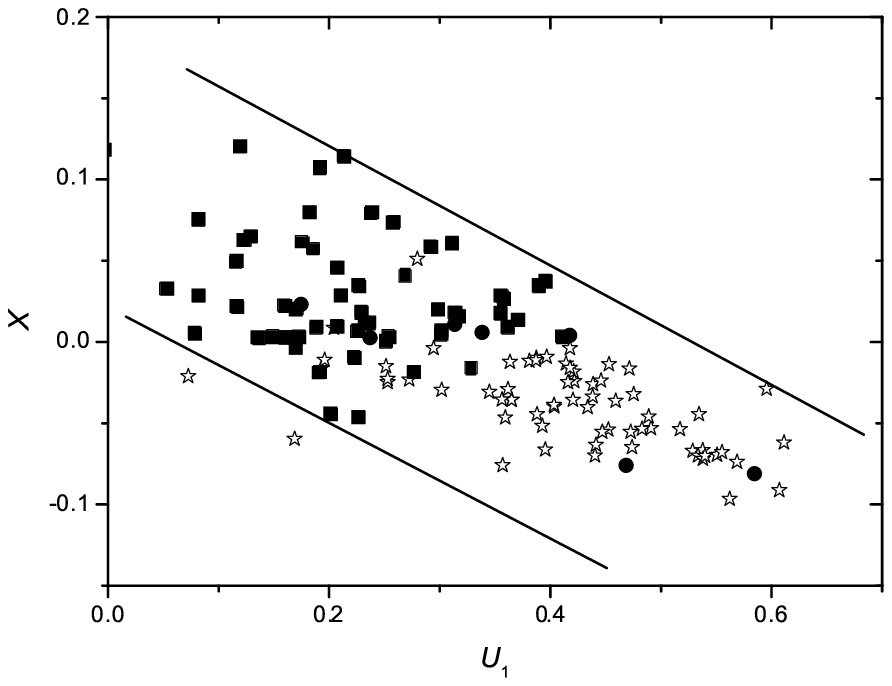,width=7.5cm}} \caption{Relations
between the temperature difference and the relative energy transfer
rate $U_2$ for A- and W-type W UMa contact binaries. The symbols are
the same as Figure 1.}
\label{fig6}
\end{figure}

The temperature difference is also an important parameter for
investigating the classification and the energy transfer of W UMa
contact binaries. It is defined by \citet{Rucinski 1974} as
\begin{equation}
{X}=\frac{T_{\rm 1}-T_{\rm2}}{T_{\rm 1}}.
\end{equation}
The temperatures of W UMa binary components can differ
substantially, and this difference was explained by different
proportions of the energy exchange in superadiabatic and adiabatic
part of envelopes \citep{Mochnacki 1973}. \citet{Rucinski 1974}
investigated the relations between the temperature difference and
other observational parameters, and found that the temperature
difference for W-type systems is not correlated with the mass ratio,
the fill-out parameter or the color.

The relation between the relative energy transfer rate ($U_1$) and
the temperature difference is shown in Figure 5. W UMa contact
binaries seem to populate a strip limited by two solid lines, and
there is a tendency for decreasing temperature difference with
increasing relative energy transfer rate. This suggests that the
temperature difference is correlated with the relative energy
transfer rate ($U_1$). This also indicates that the temperature of
the secondary increases with increasing relative energy transfer
rate and even exceeds the temperature of the primary if the relative
energy transfer rate ($U_1$) is large enough, and that a thermal
coupling exists in the two components of W UMa contact binaries.

\begin{figure}
\centerline{\psfig{figure=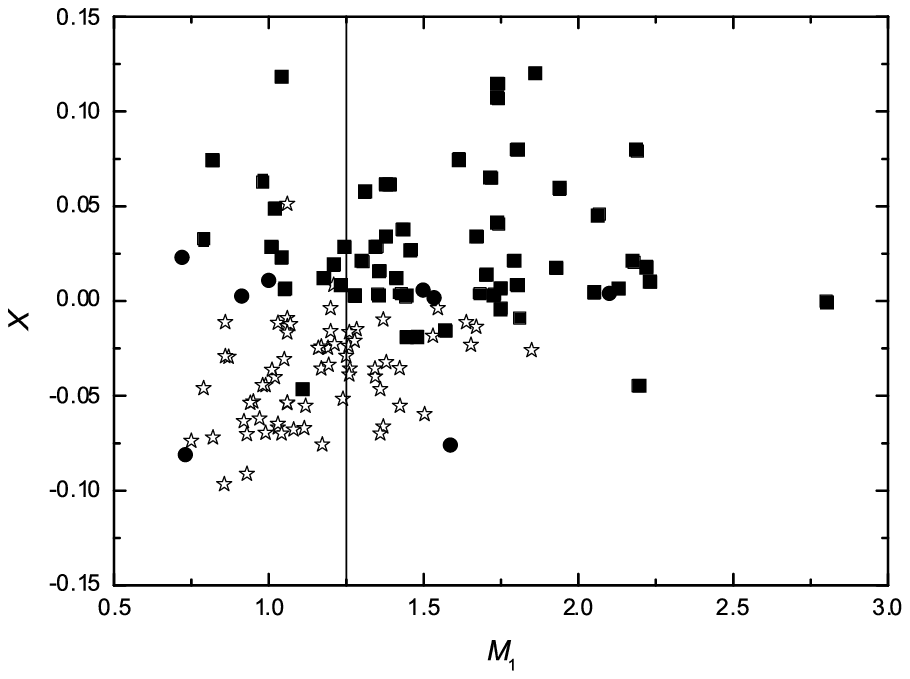,width=7.5cm}} \caption{The
distribution of temperature deviation $vs$ the mass of primary for W
UMa contact binaries. The Symbols are the same as Figure 1 and $M_1$
is in solar mass.} \label{fig7}
\end{figure}

Main-sequence stars with $M\ga1.25 M_{\rm \odot}$ have little or no
convective envelope; however, main-sequence stars with $0.35\la
M\la1.25 M_{\rm\odot}$ have a convective envelope and a radiative
core \citep{Hurley 2000}. \citet{Li 2004} argued that convection is
by no means essential to heat transport in the common envelope of W
UMa systems by employing Eggleton's stellar evolution code
\citep{Eggleton 1971, Eggleton 1972, Eggleton 1973}. This means that
the temperature difference should not depend on the mass of the
primary. The distribution of the temperature difference $X$ $vs$ the
primary's mass $M_1$ is shown in Figure 6. As seen from Figure 6,
the distribution of temperature difference of the observed systems
with $M_1\ga1.25 M_{\rm \odot}$ is similar to that of the systems
with $M_1\la1.25 M_{\rm \odot}$. This suggests that there is no
correlation between the temperature difference and the mass of the
primary, and that the efficiency of energy transfer in the common
envelope of W UMa systems is indeed not significantly enhanced by
convective motion.

\section{The effect of energy transfer on the secondaries}

\begin{figure}
\centerline{\psfig{figure=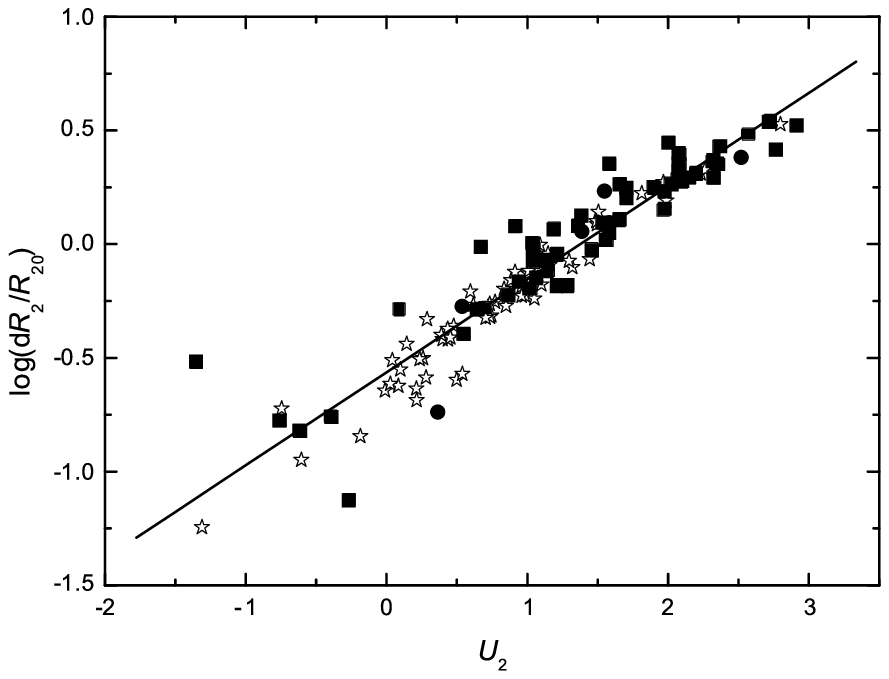,width=7.5cm}} \caption{The
relations between the relative radius change of the secondary and
the relative energy transfer rate $U_2$ for A- and W-type W UMa
contact binaries, and the solid line represents the linear fit.
Symbols are the same as Fig.1.} \label{fig8}
\end{figure}

The mass-radius relations of the secondaries for A- and W-type of
W UMa contact binaries are different from that of ZAMS stars
\citep{Yang 2001,Awadalla 2005,Li 2008}. This is a result of the
energy transfer from the primary to the secondary \citep{Webbink
2003, Li 2008}. But the relation between the energy transfer and
the radius of the secondary is not clear.

The relation between the relative radius change of the secondary
{\bf(${\rm log(d}R_2/R_{20})={\rm log}((R_2-R_{20})/R_{20})$)} and
the relative energy transfer rate ($U_2$) of W UMa contact binaries
is plotted in Figure 7. $R_{20}$ is the radius of the main sequence
secondary in W UMa systems without the effect of the energy
transfer, i.e. it is the radius of {\bf ZAMS with} mass $M_2$.
According to \citet{Lacy 1977}, it can be expressed as

\begin{equation}
R_{20}=\left\{ \begin{array}{ll}
                    0.955M_2^{0.917}  & \;\;\;\;\;\;\;\;\;\; \mbox{ $0.1\leq M_2\leq 1.318 $,} \\
            1.026M_2^{0.640}  & \;\;\;\;\;\; \mbox{ $1.318\leq M_2\leq 19.953 $,}
           \end{array}
       \right.
\end{equation}
where $M_2$ is the mass of the secondary in solar units. As seen in
Figure 7, the relative radius change of the secondary is correlated
with the relative energy transfer rate $U_2$, i.e. {\bf the radius
of the secondary in a W UMa system increases with increasing $U_2$}.
This indicates that the energy transfer exercises a decisive
influence on the secondaries of A-type and W-types. By using a
linear fitting, the relation between the relative radius change of
the secondary (${\rm logd}R_2/R_{20}$) and the relative energy
transfer rate ($U_2$) can be written as

\begin{equation}
{\rm log}({\rm d}R_2/R_{20}) =0.41(1)U_2-0.56(2).
\end{equation}
This correlation is a result of the response of the secondary to the
energy transfer from the primary. It is the energy transfer that
makes the radii of the secondaries of W UMa contact binaries deviate
from those of ZAMS stars \citep{Yakut 2005,Li 2008}. However, a W
UMa system VW Cep has a radius smaller than that given by equation
(10). VW Cep is one of triple systems \citep{Pribulla 2006}. So the
smaller secondary of VW Cep might be attributed to an inaccurate
spectroscopic solution due to the presence of the additional
companion.

\begin{figure}
\centerline{\psfig{figure=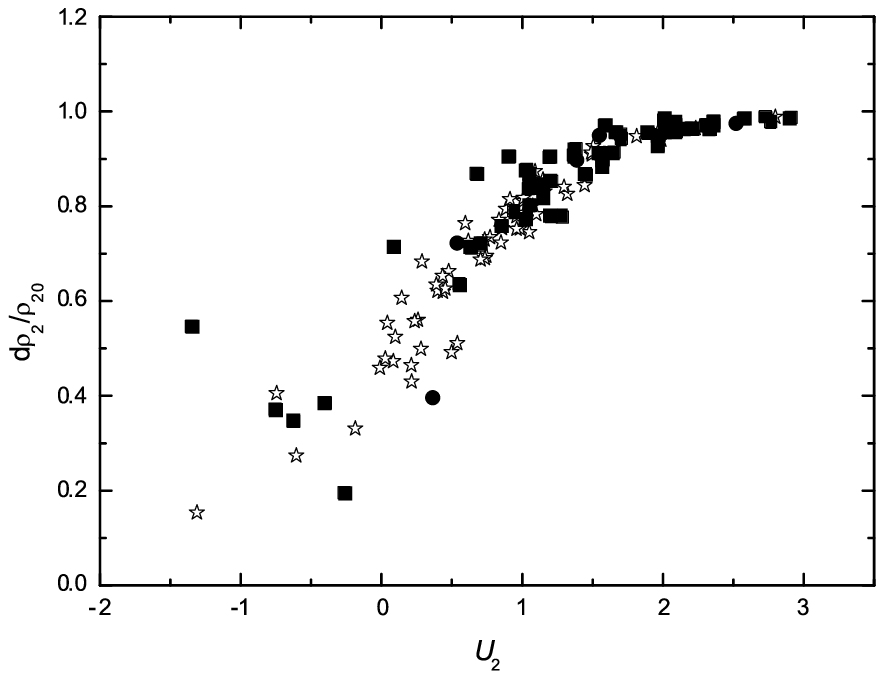,width=7.5cm}} \caption{The
relation between the relative density change of the secondary and
the relative energy transfer rate $U_2$ for A- and W-type W UMa
contact binaries. Symbols are the same as Fig. 1.}
 \label{fig9}
\end{figure}

\citet{Yang 2001} argued that the over-luminosity of the secondary
is related to its density. This is also due to the effect of the
energy transfer from the primary to the secondary on the density of
the secondary. The relation between the relative density change of
the secondary {\bf (${\rm
d}\rho_{_{2}}/\rho_{_{20}}=(\rho_{_{20}}-\rho_{_2)}/\rho_{_{20}}$)}
and the relative energy transfer rate ($U_2$) is presented in Figure
8. As shown from Figure 8, the relative density change increases
with increasing relative energy transfer rate. This suggests that
the more the secondary gets energy from the primary, the lower the
density of the secondary becomes. A secondary obtaining more energy
from the primary would swell more greatly, and then its density
would become smaller.

\section{Discussion and conclusions}

In this paper, we investigate {\bf the energy transfer }of W UMa
contact binaries based on a sample of 133 W UMa contact binaries,
and then we study the effects of the energy transfer on the
secondaries of W UMa systems.

{\bf Based on the assumptions(the components are contact
configurations with nearly uniform effective temperature and the
primaries are ZAMS), the relations are given between the relative
energy transfer rates and the mass ratio of W UMa systems. The
theoretical curves can reflect the distribution of $U_1$ $vs$ $q$
and $U_2$ $vs$ $q$. But some observation systems are significantly
deviated from these curves. By comparing the observational data and
these resulting curves, it is found that the deviations are mainly
resulted from the difference in the effective temperatures of the
components in W UMa systems. This means that the assumption that the
components are uniform in effective temperature should be
restrainedly applied to investigate the energy transfer in W UMa
systems.}

The distribution of the temperature difference $vs$ the mass of the
primary suggests that the convection does not affect the efficient
of energy transfer between two components in the common envelope of
W UMa contact binaries. This suggests that the energy transfer in W
UMa systems does not depend on the property of the common envelope
of W UMa contact binaries and the convection is by no means
essential to heat transport in the common envelope of W UMa systems.
This also suggests that the energy transfer might occur in radiative
region of common envelope of W UMa contact binaries \citep{Li 2004}
or the mechanism of energy transfer might be the differential
rotation \citep{Yakut 2005} or circulation currents \citep[][and
references therein ]{web77,rob80}.

The energy transfer from the primary to the secondary would lead W
UMa systems to be not in thermal equilibrium, then lead W UMa
contact binaries to suffer thermal relaxation oscillations (TRO).
However, the energy transfer is also related to the evolutionary
degree of the primary, i.e. the higher is the evolutionary degree of
the primary, the lower is the energy transfer rate \citep{Liu 2000}.
This suggests that with the evolution of W UMa systems, the thermal
relaxation oscillation might be disappeared if the evolutionary
degree of the primary is high enough. If the energy transfer rate
decreases in the evolved W UMa systems, the rate of mass transferred
from the secondary to the primary should become smaller and smaller
with the evolution of W UMa systems, and the decrease in mass ratio
of the systems would become slower and slower, so that the lifetime
of W UMa systems might become longer than the prediction of the
theory models \citep{Li 2005}.

Figure 7 shows that the relative radius change of the secondary
increases with increasing relative energy transfer rate $U_2$.
\citet{Webbink 2003} and \citet{Li 2008} suggested that the the
deviation of the radius of the secondary from that of ZAMS stars is
probably the result of the energy transfer from the primary to the
secondary in W UMa contact binaries. The energy transfer from the
primary to the secondary is more than the energy generated in the
core of the secondary, so the radius and density of the secondaries
is significantly influenced by the energy transfer and their radius
and density deviates from those of ZAMS stars. \citet{Hazlehurst
1977} calculated the effects of energy transfer on the radius and
temperature of stars and gave the response functions for the radius
and temperature of the stars. In present paper, we have given a
relation between the relative radius change of the secondary and the
relative energy transfer rate $U_2$. This relation probably provides
a useful information in the structure of the secondaries in W UMa
contact binaries, and can help us to understand the structure and
evolution of W UMa systems.

\section*{ACKNOWLEDGEMENTS}

The authors are grateful to an anonymous referee for his/her
valuable suggestions and insightful remarks, which have improved
this paper greatly. This work was partly supported by the Chinese
Natural Science Foundation (10673029, 10773026, 10433030 and
10521001), and by the Yunnan Natural Science Foundation (2007A113M
and 2005A0035Q).

\bsp

\label{lastpage}

\end{document}